\documentclass[preprint,nofootinbib,tightenlines]{revtex4}
\usepackage{amsmath}

\usepackage{bm}
\usepackage{braket}

\makeatother
\begin{document}
\title{Schiff screening of relativistic nucleon electric-dipole moments by electrons}

\author{C.-P. Liu}
\email{cpliu@lanl.gov} \affiliation{T-16, Theoretical Division, Los Alamos National Laboratory,
Los Alamos, NM, USA 87545}
\author{J. Engel}
\email{engelj@physics.unc.edu} \affiliation{Department of Physics
and Astronomy, University of North Carolina, Chapel Hill, NC, USA
27599-3255}
\preprint{LA-UR-07-2423}

\begin{abstract}
We show, at leading-order in the multipole expansion of the electron--nucleus
interaction, that nucleon electric-dipole moments are completely
shielded by electrons so that they contribute nothing to atomic
electric-dipole moments, even when relativity in the nucleus is
taken into account. It is well known that relativistic electron
motion, by contrast, leads to dipole moments that are not screened;
we discuss the reasons for the difference.
\end{abstract}
\maketitle

Some years ago, Schiff~\cite{Sch} showed that in the limit of a
pointlike nucleus and nonrelativistic electrons, any electric-dipole
moments (EDMs) carried by the electrons and nucleus are completely
screened by atomic polarization, so that the EDM of the atom
vanishes.  Shortly thereafter, Sandars~\cite{San} pointed out that
in heavy polarizable atoms, relativistic corrections to the
electron-EDM operator not only survive screening but can enhance
atomic EDMs. For nucleons in a nucleus, $v/c$ is about 0.1, and one
may wonder whether relativity allows nucleon EDMs to evade screening
at some level.  Ref.~\cite{OFA} claims that it does, at a level much
larger than that resulting from the finite size of the nucleus.  The
authors use the result to argue that the limit on the neutron EDM
from experiments in $^{199}$Hg should be $|d_n| \ \raisebox{-.25ex}{$\stackrel{<}{\scriptstyle \sim}$}\ 2.5 \times
10^{-26} e\,\mathrm{cm}$, a value nearly an order of magnitude
tighter than the generally accepted limit, $|d_n| \raisebox{-.25ex}{$\stackrel{<}{\scriptstyle \sim}$}\  4 \times 10^{-25}
e\,\mathrm{cm}$~\cite{DS}, and comparable to the current best limit
from direct measurement, $|d_n| < 2.9 \times 10^{-26}
e\,\mathrm{cm}$~\cite{Bak}.   But nucleons are different from
electrons in that they are confined to a much smaller volume. This
confinement, as we show below, implies that relativistic
contributions to nucleon EDMs are just as screened as their
nonrelativistic counterparts.

To see this, let us divide the parity-and-time-reversal-conserving
part of the system's internal Hamiltonian $H_0$ into a piece that
involves electrons only --- the electron kinetic energy and
electron-electron interactions --- a corresponding part that
involves nucleons only, and a part that contains the
electron-nucleon interaction:
\begin{align}
\label{eq:H_0} H_{0} &
=H_{e}^\mathrm{int}+H_{\mathrm{nuc}}^\mathrm{int}+H_{e\textrm{--}\mathrm{nuc}}\,,
\end{align}
where ``int'' stands for ``internal''. We expand electron--nucleus
part of $H_0$ (which is the same Hamiltonian as the $H_0$ in
Eq.~(2.3) of Ref~\cite{OFA}) in multipoles:
\begin{align}
\label{eq:eNCoul}
H_{e\textrm{--}\mathrm{nuc}} & =\sum_{i=1}^{Z}\,\frac{Z\,
e^{2}}{r_{i}}+\ldots\,,
\end{align}
where $r_{i}\equiv|\bm r_{i}|$ is the coordinate of the
$i^{\mathrm{th}}$ electron. We shall neglect in this discussion the
sub-leading terms {}``$\ldots$'', which involve nuclear moments
(static and local) beyond the lowest-order monopole. These higher moments,
corresponding to what are often called ``finite-size effects", do in fact allow Schiff
screening to be evaded, but at a low level that is
systematically discussed in Ref.~\cite{LR-MHTD}.

We can now calculate corrections to the ground-state energy induced by the
parity- and time-reversal-violating interaction of the nucleon dipole moments with an external electric field $\bm
E_\mathrm{ext}$.  The first order shift is
\begin{align}
\Delta E^{(1)} & =-\langle \mathrm{g.s.}| \ \sum_{j=1}^{A}\, d_{N}^{j}\,\gamma^0_{j}\,
\bm\Sigma_{j} \ |\mathrm{g.s.}\rangle\cdot\bm E_{\mathrm{ext}}\,,
\label{eq:dE1}
\end{align}
where $\gamma^0_{j}=\left(\begin{array}{cc}
1 & 0\\
0 & -1\end{array}\right)$ is the Dirac matrix for the $j^{\mathrm{th}}$ nucleon,
$\bm\Sigma_j =  \left(\begin{array}{cc}
\bm\sigma_j & 0\\
0 & \bm\sigma_j \end{array}\right)$ is the spin operator for that
nucleon, $d_{N}^{j}$ is the magnitude of the EDM for the same
nucleon, and $|\mathrm{g.s.}\rangle$ is the ground state of the
unperturbed Hamiltonian $H_0$. Often the relativistic operator in Eq.~(\ref{eq:dE1}) is
divided into two pieces by writing $\gamma^0$ as $1+(\gamma^0-1)$;
the second piece is then purely relativistic (see, {\it e.g.}, Ref.\
\cite{San}). We do not make this split because for nucleons both
pieces will be shielded.

To see that no portion of the dipole operator is left unscreened in
the point-like (lowest-order monopole) limit, we now calculate the second-order
addition to Eq.~(\ref{eq:dE1}) from polarization of the electrons.
With the nucleus kept in its ground state (see below for justification), this contribution is
\begin{align}
\Delta E^{(2)}= &
-\sum_{n}\,\frac{1}{E_{\mathrm{g.s.}}-E_{n}}\,\braket{\mathrm{g.s.}|\left(\sum_{j=1}^{A}\,
d_{N}^{j}\,\gamma^0_{j}\,\bm\Sigma_{j}\right)\cdot\left(\sum_{i=1}^{Z}\,\bm\nabla_i
A_{0}\right)|n}
\nonumber \\
 & \times\braket{n|e\,\sum_{k=1}^{Z}\,\bm r_{k}\cdot\bm E_\mathrm{ext}|\mathrm{g.s.}}+\mathrm{H.c.}\,,
\label{eq:dE2}
\end{align}
where the label $n$ on the states refers to electronic
configurations, and $A_{0}$ denotes the electric potential at the
origin (where the nucleus is located) generated by the electrons:
\begin{align}
A_{0} & = \sum_{i=1}^Z\,\frac{e}{r_{i}}=-\frac{1}{Z\, e}\,
H_{e\textrm{--}\mathrm{nuc}}\,.
\label{eq:E_int}
\end{align}
The gradient $\bm\nabla_i\,A_0$ in Eq.~(\ref{eq:dE2}) is then just
the electric field at the origin produced by the $i^\mathrm{th}$
electron.  In writing these expressions we have assumed only that the full wave
function factors into products of atomic and nuclear wave functions to
good approximation.

The truncation of the multipole expansion at leading order ---
the point-like approximation ---  allows us to restrict
attention to the electric field at the origin and write
$H_{e\textrm{--}\mathrm{nuc}}$ in terms of electron operators only in
Eq.~(\ref{eq:E_int}).
As a consequence, the nuclear state is not perturbed; excited
nuclear states do not contribute to Eq.~(\ref{eq:dE2}), even if the
coupling of nucleons to $\bm E_\mathrm{ext}$ is included (nuclear
polarization).  For there to be a contribution, the excitations
would have to be created by a nuclear operator with positive parity
({\it i.e.}, $\bm \Sigma$) and destroyed by one with negative parity
({\it i.e.}, $\bm r$). That can't happen, however, because
$H_\mathrm{nuc}^\mathrm{int}$ is symmetric under reflection so that all
unperturbed nuclear states have good parity.

These facts allow us evaluate sum in $\Delta E^{(2)}$.
From Eq.~(\ref{eq:E_int}) it follows that
\begin{align}
\left(\sum_{j=1}^{A}d_{N}^{j}\,\gamma^0_{j}\,\bm\Sigma_{j}\right)\cdot\left(\sum_{i=1}^{Z}\,\bm
\nabla_i A_{0}\right) & =-\frac{1}{Z\,
e}\,\left[\sum_{j=1}^{A}\,\sum_{i=1}^{Z}\,
d_{N}^{j}\,\gamma^0_{j}\,\bm\Sigma_{j}\cdot\bm\nabla_{i}\,,\,
H_{e\textrm{--}\mathrm{nuc}}\right]
\nonumber \\
 & =-\frac{1}{Z\, e}\,\left[\sum_{j=1}^{A}\,\sum_{i=1}^{Z}\,
d_{N}^{j}\,\gamma^0_{j}\,\bm\Sigma_{j}\cdot\bm\nabla_{i}\,,\, H_{0}-H_{e}^{\mathrm{int}}-
H_{\mathrm{nuc}}^{\mathrm{int}}\right]\,.
\label{eq:commutator}
\end{align}
Now, noting that
\begin{enumerate}
\item
the electron-electron interaction is pair-wise, so that
\begin{align}
\sum_{i=1}^{Z}\,[d_{N}^{j}\,\gamma^0_{j}\,\bm\Sigma_{j}\cdot\bm\nabla_{i}\,,\,
H_{e}^{\mathrm{int}}] & = \sum_{i=1}^{Z}\,d_{N}^{j}\,\gamma^0_{j}\,\bm\Sigma_{j}\cdot[\bm\nabla_{i}\,,\,
H_{e}^{\mathrm{int}}]=0\,,
\label{eq:d-He}
\end{align}
and
\item the nucleus is not excited so that the commutator of any nuclear operator with
$H_\mathrm{nuc}$ yields a vanishing expectation value:
\begin{align}
_{\mathrm{nuc}}\braket{\mathrm{g.s.}|[d_{N}^{j}\,\gamma^0_{j}\,
\bm\Sigma_{j}\cdot\bm\nabla_{i}\,,
\,H_{\mathrm{nuc}}^{\mathrm{int}}]|\mathrm{g.s.}}_{\mathrm{nuc}} &
={}_{\mathrm{nuc}}\braket{\mathrm{g.s.}|[d_{N}^{j}\,\gamma^0_{j}\,\bm\Sigma_{j}\,,
\,H_{\mathrm{nuc}}^{\mathrm{int}}]|\mathrm{g.s.}}_{\mathrm{nuc}}\cdot\bm\nabla_{i}=0\,,
\label{eq:d-Hnu}
\end{align}
\end{enumerate}
we can eliminate the energy denominators in $\Delta E^{(2)}$ and
perform the sum over intermediate state in closure. The result is
\begin{align}
\Delta E^{(2)}= &
\frac{1}{Z}\,\braket{\mathrm{g.s}|\left[\left(\sum_{j=1}^{A}\,\sum_{i=1}^{Z}\,
d_{N}^{j}\,\gamma^0_{j}\,\bm\Sigma_{j}\cdot\bm\nabla_{i}\right),\,\sum_{k=1}^{Z}\,\bm r_{k}
\cdot\bm E_{ext}\right]|\mathrm{g.s.}}\nonumber \\
= & \langle \mathrm{g.s.}| \ \sum_{j=1}^{A}\,
d_{N}^{j}\,\gamma^0_{j}\,\bm\Sigma_{j} \ |\mathrm{g.s.}\rangle \cdot
\bm E_{\mathrm{ext}}=-\Delta E^{(1)}\,,
\label{eq:screening}
\end{align}
so that the first- and second-order contributions cancel each other
exactly. Thus, atomic polarization screens the nucleon EDMs even in
relativistic quantum theory.  The finite size of the nucleus,
leading to a difference between the monopole and dipole charge
densities, is still the dominant nuclear contribution to the atomic
EDM. Relativistic corrections to nuclear wave functions affect the
result only a little because their contributions to densities are of
$\mathcal{O}(v^2/c^2) \approx 1\%$.

Why then do relativistic corrections to \emph{electron} EDM
operators evade screening?  The reason is that the $\gamma^0$ in the
relativistic operator doesn't commute with the relativistic
free-electron Hamiltonian $\gamma_0 (m_e + \bm p \cdot \bm \gamma
)$, so that the first commutator in Eq.~(\ref{eq:d-He}) doesn't
vanish if the Dirac matrices act on electrons (see
Refs.~\cite{San,LLS,LR-MHTD} for details). By contrast,
Eq.~(\ref{eq:d-Hnu}) vanishes even if one uses a relativistic form
for the nuclear Hamiltonian because of the expectation value. And as
mentioned above, off-diagonal nuclear matrix elements contribute
nothing because of the parity symmetry of
$H_\mathrm{nuc}^\mathrm{int}$.

In summary, at leading-order in the multipole expansion for
$H_{e\textrm{--}\mathrm{nuc}}$, where the electrons see the
nucleus as a point particle, even fully relativistic nucleon EDMs
are screened by electron polarization. The effects of relativity in
the nucleus will only add small corrections to the usual finite-size
effects encoded in the nuclear ``Schiff moment''.

\begin{acknowledgments}
This work was supported in part by the U.S. Department of Energy under Contracts DE-AC52-06NA25396 (CPL) and DE-FG02-97ER41019 (JE).
\end{acknowledgments}


\begin{thebibliography}{1}
\bibitem{Sch} L. I. Schiff, Phys. Rev. 132, 2194 (1963).
\bibitem{San}P. G. H. Sandars, J. Phys. B \textbf{1}, 511 (1968).
\bibitem{OFA}Sachiko Oshima, Takehisa Fujita, and Tomoko Asaga, Phys. Rev. C \textbf{75}, 035501 (2007).
\bibitem{DS}V. F. Dmitriev and R. A. Sen'kov, Phys. Rev. Lett. \textbf{91}, 212303 (2003).
\bibitem{Bak}C. A. Baker {\it et al.}, Phys. Rev. Lett. \textbf{97}, 131801 (2006).
\bibitem{LR-MHTD}C.-P. Liu, M. J. Ramsey-Musolf, W. C. Haxton, R. G. E. Timmermans, and A. E. L. Dieperink, arXiv:0705.1681 [nucl-th].
\bibitem{LLS}E. Lindroth, B. W. Lynn, and P. G. H. Sandars, J. Phys. B \textbf{22}, 559 (1989).
\end{thebibliography}
\end{document}